\documentclass[runningheads]{llncs}
\usepackage[T1]{fontenc}
\usepackage{cite}
\usepackage{amsmath,amssymb,amsfonts}
\usepackage{algorithmic}
\usepackage{textcomp}
\usepackage{booktabs}
\usepackage{arydshln}
\usepackage{subfig}
\usepackage{siunitx}
\usepackage{caption}
\usepackage{multirow}
 \usepackage{xspace}
 \usepackage[utf8]{inputenc}
\usepackage[table,svgnames]{xcolor} 

\usepackage{graphicx}
\begin{document}

\title{Smart Video Capsule Endoscopy: Raw Image-Based Localization for Enhanced GI Tract Investigation\thanks{This work has been partly funded by the German Federal Ministry of Research, Technology and Space (BMFTR) in the projects MEDGE (16ME0530) and Scale4Edge (16ME012).}}
\titlerunning{Smart Video Capsule Endoscopy: Raw Image-Based Localization}
%
\author{Oliver Bause\orcidID{0009-0003-5388-2959}† \and
Julia Werner\orcidID{0009-0006-0279-1776}† \and 
Paul Palomero Bernardo\orcidID{0000-0002-6642-3976} \and
Oliver Bringmann\orcidID{0000-0002-1615-507X}}
\authorrunning{Bause and Werner et al.}
%
\institute{Chair of Embedded Systems, Eberhard Karls University of Tübingen,\\72076 Tübingen, Germany\\
\email{oliver.bause@uni-tuebingen.de\\julia-helga.werner@uni-tuebingen.de}\\
\url{https://www.embedded.uni-tuebingen.de/en/home/}
\def\thefootnote{†}\footnotetext{Equal contribution: These authors contributed equally to this work.}}
\maketitle              
\begin{abstract}
For many real-world applications involving low-power sensor edge devices deep neural networks used for image classification might not be suitable.
This is due to their typically large model size and requirement of operations often exceeding the capabilities of such resource limited devices.
Furthermore, camera sensors usually capture images with a Bayer color filter applied, which are subsequently converted to RGB images that are commonly used for neural network training.
However, on resource-constrained devices, such conversions demands their share of energy and optimally should be skipped if possible.
This work addresses the need for hardware-suitable AI targeting sensor edge devices by means of the Video Capsule Endoscopy, an important medical procedure for the investigation of the small intestine, which is strongly limited by its battery lifetime.
Accurate organ classification is performed with a final accuracy of $93.06\%$ evaluated directly on Bayer images involving a CNN with only 63,000 parameters and time-series analysis in the form of Viterbi decoding.
Finally, the process of capturing images with a camera and raw image processing is demonstrated with a customized PULPissimo System-on-Chip with a RISC-V core and an ultra-low power hardware accelerator providing an energy-efficient AI-based image classification approach requiring just \qty{5.31}{\micro\joule} per image. 
As a result, it is possible to save an average of $89.9\%$ of energy before entering the small intestine compared to classic video capsules.

\end{abstract}

\section{Introduction}
Vision-based deep neural networks are commonly trained with standard RGB images.
While cameras typically capture images in Bayer pattern format, they are normally converted to RGB images afterwards for further utilization.
In the field of machine learning, this circumstance is often not addressed since, for many applications, conducting such a conversion can be easily performed.
However, for a range of real-world applications relying on small, low power edge devices this conversion is costly and should be avoided to save the limited power if possible.\\
\hspace*{2em}Video Capsule Endoscopy (VCE) is an example for a medical application that requires an edge device with limited power availability~\cite{iddan2000wireless,swain2001wireless}.
This medical procedure was first introduced in the early 2000s and comprises the swallowing of a small pill-sized capsule with LEDs and a camera that can capture images of the digestive tract while the capsule moves from the esophagus through the stomach and small intestine to the colon.
Subsequently, the images are transmitted to an external on-body receiver for further inspection by qualified physicians.
This examination of the gastrointestinal (GI) tract is crucial for the investigation of the otherwise largely inaccessible small intestine which can exhibit gastroenterological pathologies such as ulcers, inflammation, polyps or cancer.
Due to the limited volume of the capsule, the overall battery lifetime is severely restricted.
For example, for the PillCam™ SB3 from Medtronic an operation time of \qtyrange[range-units = single]{8}{12}{\hour} is reported with \qtyrange[range-units = single]{2}{6} frames per second (fps)~\cite{pillcam,monteiro2016pillcam}.
However, the time of passing through the whole GI tract varies from patient to patient and can easily exceed \qty{12}{\hour} in certain cases.
Given the human anatomy of the GI tract, this can result in an incomplete screening which is especially problematic if anomalies or diseases are located in the uncovered region of interest.\\
\hspace*{2em}\textbf{Our Contribution: } 
To prolong the battery lifetime of small edge devices, that incorporate artificial intelligence (AI) for image classification, this work aims to reduce the energy consumed per inference by using raw Bayer instead of the commonly used RGB images.
This enables a precise localization of the capsule within the GI tract to determine the moment when the region of interest, the small intestine, is entered. 
Images obtained from the preceding organs are discarded and the transmission only starts after this time point. 
Furthermore, the frame rate before entering the small intestine can be reduced which also saves power.
Moreover, first experiments on training a very light-weight convolutional neural network (CNN) directly on images in Bayer pattern format are conducted which saves an additional conversion step on hardware and also provides a baseline for future experiments.
The whole process is demonstrated using a customized PULPissimo System-on-Chip (SoC) with a RISC-V core and an ultra-low power hardware accelerator.
This process consits of capturing VCE images with a miniature camera, analyzing them with a small CNN, and parsing the labels to a microcontroller unit (MCU) for post-processing with Viterbi decoding.
To conclude, we provide an energy-efficient AI classification pipeline for vision applications and demonstrate this with the example of VCE.
\section{Related Work}

\subsection{CNN Training on Bayer images}
Mobile devices are generally equipped with camera sensors that capture images with a Bayer color filter applied.
To improve the quality of these images, it is necessary to demosaic and denoise them.
Research is ongoing to enhance the post-processing by leveraging neural networks~\cite{khadidos2021bayer,kumar2023noising}.
However, the integration of such networks into ultra-low power devices is not a viable solution, as these are characterized by high computational demands.
Additionally, CNNs employed for vision tasks are typically trained on three-channel RGB images rather than raw, mosaicked Bayer images.
\cite{chandra2021novel} investigated the potential of training SqueezeNet with Bayer images to classify hand postures.
The accuracy of the method was determined to be $98.28\%$, with a co-sited demosaicing technique employed.
However, when using raw Bayer images, the accomplished accuracy was less than $25\%$.

\subsection{Hardware-Suitable CNNs for Video Capsule Endoscopies}

Precise organ classification based on VCE images has been substantially progressed due to the release of a publicly available VCE dataset, the Rhode Island Gastroenterology dataset~\cite{charoen2022rhode}, with annotations for the four main organs of the GI tract, which are traversed by a video capsule.
\cite{charoen2022rhode} provides a baseline with an Inception ResNetV2, achieving an accuracy of \qty{97.1}{\percent}.
This problem was further addressed by \cite{werner2023precise} and \cite{abian2025atrous}, yielding accuracies of \qty{96.95}{\percent} and \qty{69.84}{\percent}, respectively.
While \cite{werner2023precise} aimed to lower the model complexity and constrain the total number of parameters to 1M, the proposed approaches still demand too many resources from potential hardware architectures.
\cite{palomero2025compiler} provided first results on a programmable edge AI accelerator which can efficiently execute a simple convolutional network while demonstrating proficient results on the VCE task. 
Importantly, all those findings are based on CNN evaluations using RGB images.
~\cite{sahafi2022edge} reported first results on real-time VCE image processing with onboard neural networks.
However, the neural network employed, still consisted of $\approx 3$ M parameters and their prototype did not exceed an operation time of \qty{1}{\hour}, limiting the applicability due to a high power consumption.

Based on this previous research, we aim to perform neural network training directly on images in Bayer pattern format to omit the conversion step to the RGB format combine the CNN evaluations with post-processing with an HMM and Viterbi decoding as shown by~\cite{werner2023precise} with additional quantization and finally demonstrate this with a hardware-in-the-loop (HIL) testing~\cite{bause2025hil}.

\section{Methodology}

\begin{figure*}[hptb]
    \centering
    \includegraphics[width=\linewidth]{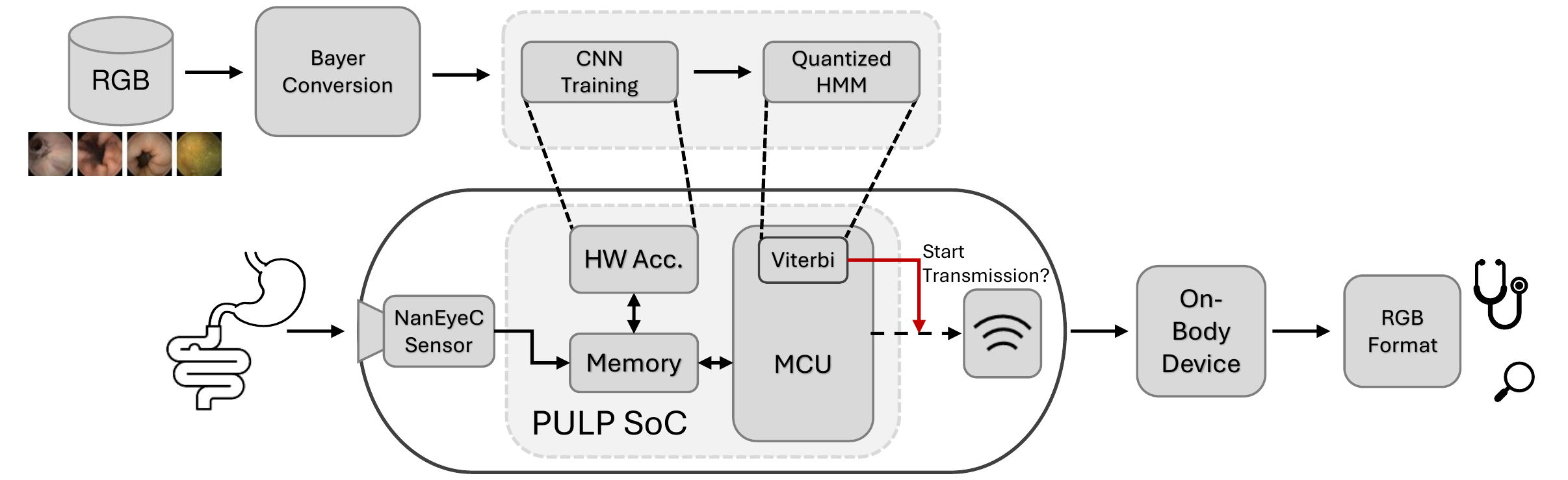}
    \caption{Proposed method of processing raw input VCE images with a quantized low-complex CNN and quantized Viterbi decoding on a VCE demonstrator involving the NaneyeC camera, a hardware accelerator and an MCU.}
    \label{fig:method}
\end{figure*}

The proposed method is illustrated in Figure~\ref{fig:method} involving the training of a quantized CNN which receives VCE images in raw Bayer pattern format as an input and parses its evaluations to a quantized HMM. 
Since the camera forwards images directly in Bayer pattern format and conversion to RGB images requires it's share of energy and memory, the dataset at hand is converted to Bayer pattern frames to simulate the conditions of this medical application realistically.
Additionally, the capsule's composition is simulated with an asm NanEyeC miniature camera that sends images in Bayer pattern format to a hardware accelerator that can efficiently execute the CNN and features an MCU for subsequent processing of quantized Viterbi decoding.
During the application, this decision pipeline should help to decide, when to start transmission of VCE frames to an on-body device and adjust the frame rate to lower the overall power consumption. Finally, only when outside of the capsule, the images are converted to RGB format for evaluations by medical experts.

\subsection{Bayer Filter Conversion}
Image sensors, especially those with a miniature form factor, capture images with a color filter array (CFA) applied~\cite{bayer1976,naneyec}.
The Bayer filter is the most common one and uses a $2\times2$ pattern with two green, one red, and one blue pixel. 
The resulting filters are called BGGR, RGBG, GRGB, and RGGB.
Thus, Bayer images are space-saving as they only have one color channel per pixel.
To obtain an RGB image from a raw Bayer image, the missing color channels of each pixel need to be calculated by interpolating the corresponding color channels of the neighboring pixels.
Further demosaicing can then be performed to increase image quality.

Images within datasets are normally stored in an RGB image format like PNG or JPEG.
Nevertheless, these were often captured using a sensor with a CFA. 
Consequently, it is feasible to convert them back into the original format with minimal loss of data by extracting the original color filter again.

\subsection{Rhode Island Gastroenterology Dataset}
Training neural networks for performing localization within the GI tract based on VCE images requires an extensive database of VCE studies.
The Rhode Island gastroenterology dataset~\cite{charoen2022rhode} is one of the largest publicly available VCE datasets and consists of  $5,247,588$ labeled images from $424$ VCE studies in total.
They feature annotations for the four organs: esophagus, stomach, small intestine and colon.
In this work, this dataset along with its official splits of training, validation and test set was adopted for all further experiments.

\subsection{CNN Architectures and Training Specifications}
Neural networks from the Mobilenet~\cite{howard2019searching} family have been specifically designed to target applications for mobile devices and are thus less complex than other common neural networks (e.g. large ResNets or vision transformer).
Furthermore, these model architectures have been used for the problem of classifying low-resolution images from VCEs recently~\cite{werner2023precise}.
Following those findings, we first employed a MobileNetV3-Small architecture in 32-bit floating point representation which was pretrained on Imagenet to explore the classification performance of this model with standard RGB images as an input compared to raw Bayer pattern images using the same training conditions.

However, considering that this model still consists of 1M parameters in total, lowering the model complexity is still desirable. Thus, an even less complex, residual network that has been recently introduced by~\cite{palomero2025compiler}, was adapted and tested with RGB images as well as inputs in raw Bayer pattern format.
This model consists of less than $100,000$ parameters and has been successfully deployed on an AI hardware accelerator with high energy efficiency.
All models were trained for $10$ epochs with a learning rate of $0.001$ and a weight decay of $10^{-5}$ using the AdamW optimizer in PyTorch~\cite{Ansel_PyTorch_2_Faster_2024}. 

\subsection{Hidden Markov Model and Viterbi Decoding}
As shown in~\cite{werner2023precise}, the path of such video capsule can be modeled by a Hidden Markov Model (HMM) in which each organ is represented as a hidden state and the transition probabilities mark the transitioning into a new organ. The emission probabilities model the probability of observing a certain organ while actually being in a specific organ at a moment $t$. 
Subsequent Viterbi decoding~\cite{forney1973viterbi} returns the most likely path of organs based on the CNN evaluations.
We adapted this approach with few modifications to ensure hardware suitability.
As conducted by~\cite{werner2023precise}, we compute the Viterbi algorithm using solely the log likelihoods, which only require additions instead of multiplications as operations.
This prevents numerical instability in case of very small likelihoods.
However, instead of floating-point operations, we employ strict fixed point representation and quantize not only the weights, bias and features of the CNN but, additionally, generate a quantized Viterbi algorithm in Python as well as C, that can be directly transferred to the following hardware architecture.

\subsection{System Architecture}
The system architecture of our VCE demonstrator relies on a customized PULPissimo SoC which features a RISC-V core~\cite{bernardo2024scalable,schiavone2018pulpissimo}.
The chip can be clocked between \qtyrange[range-units = single]{20}{400}{\MHz} and features \qty{384}{\kibi\byte} of SRAM.
A stringent clock-gating approach which involves regulating the core, the
accelerator, and the memories enables the utilization for energy-constrained applications.

The SoC has been extended with an integrated hardware controller to efficiently communicate with the asm NanEyeC miniature camera sensor~\cite{naneyec}.
The camera captures images with a resolution of $320\times320$ pixels at up to $58$fps with a RGGB CFA applied and a size of just \qty{1}{\mm\squared}.
The controller handles the configuration and communication with the camera autonomously to off-load the RISC-V core and to save energy.
Compared to the camera's generic SPI protocol, the controller also formats and stores the received image data as packed \texttt{uint8} in the SoC's memory independently.

Our custom PULPissimo is equipped with the UltraTrail AI accelerator~\cite{bernardo2020ultratrail} for real-time, ultra-low power inference of temporal convolutional networks.
The accelerator enables the SoC to perform on-the-edge AI image classification with 64 Multiply-and-Accumulate (MAC) units.
UltraTrail directly accesses the SoC's memory banks via the TCDM interconnect to retrieve the features and weights and to store the results of the current layer.
Thus, it only requires a dedicated \qty{136}{\kibi\byte} SRAM for interim results.

The PULPissimo SoC is synthesized in \texttt{22FDX+} technology and has a chip area of only \qty{2}{\milli\metre} by \qty{1.8}{\milli\metre}.
Simulations are performed to obtain energy consumption estimations of a taped-out design that can be integrated into a VCE prototype.
However, to verify the functionality of the proposed architecture, the chip is programmed onto the Digilent Nexys A7-100T FPGA~\cite{NexysA7Digilent}.
This allows us to simulate the whole system within a controlled environment.
The FPGA is connected to the HIL setup presented in~\cite{bause2025hil}.
The HIL offers a digital twin of the NanEyeC camera sensor and is connected to a database containing the Rhode Island~\cite{charoen2022rhode} test set.
Thus, it is possible to simulate the traversal of the VCE prototype through the studies' GI tracts.
The accelerator's output is verified against our pre-trained model and the original labels of the data set.
Various capsule settings can be tested in real-time to find the optimal configuration in regard to energy consumption, frame rate and latency.
Additionally, functional correctness can be assured before performing time- and cost-intensive animal testing, reducing the overall development cycle.

The firmware executed on the PULPissimo SoC has the task of utilizing the modules and sensors of the capsule in order to achieve the best possible screening of the area of the GI tract to be analyzed.
The traversal through the entire GI tract can take up more than \qty{12}{\hour}.
Therefore, the available energy must be utilized in a carefully targeted manner.
The firmware captures images with a dynamic frame rate depending on the current system state.
The images are then analyzed by the hardware accelerator and a memory-efficient quantized implementation of the Viterbi decoding to determine the position of the capsule. 
As long as it is still not in the small intestine, the frame rate can be reduced and the transmission of the images to the on-body receiver can be deactivated to further save energy.

\section{Results}

\subsection{Classification Performance: RGB - Bayer}
All baseline studies for this dataset were conducted using RGB images.
However, as previously mentioned, converting the initial Bayer pattern format to the standard RGB image, demands its share of energy.
Thus, the question arises, how well a CNN can process and classify images in Bayer pattern format compared to RGB images.
As a first step, we compared the light-weight MobileNet trained on either RGB or Bayer pattern images in comparison to the baseline results from the literature (see Table~\ref{tab:results_mobilenet}).

\begin{table}[htbp]
	\caption{Comparison of the light-weight MobileNet trained on either RGB or Bayer pattern images (RGGB) in comparison to literature results.}
    \begin{center}\resizebox{\columnwidth}{!}{
		\begin{tabular}{lcccccc}
            \toprule
             &\textbf{Input} & \textbf{Accuracy} &  \textbf{Recall} & \textbf{Precision}  & \textbf{F1-Score} & \textbf{Params} \\
            \midrule\midrule
            Inception ResNetV2 ~\cite{charoen2022rhode} &  RGB & $97.1$ & $97.07$ & $97.31$ & $97.13$ & 56 M \\
            Swin Transformer~\cite{abian2025atrous} & RGB & $69.84$ & $69.91$ & $69.72$ & $69.85$ & 195 M \\
            MobileNetV3\cite{werner2023precise} & RGB & $96.95$ & -- & --& --& 1 M \\\hline
            MobileNetV3 & RGB & $97.14$ & $95.74$ & $87.25$ & $90.89$ & 1 M \\
            MobileNetV3 & RGGB & $96.20$ & $94.30$ & $83.92$ & $88.12$ & 1 M \\
            \bottomrule
        \end{tabular}%
        }
    \end{center}
\label{tab:results_mobilenet}
\end{table}

All previous experiments on this dataset were conducted with RGB images as an input, which functions as the standard in the field of computer vision.
Based on~\cite{werner2023precise}, we adapt the usage of the low-complex MobileNet architecture for this image classification problem achieving a similar accuracy of \qty{97}{\percent} (compared to $96.95\%$) with a total of 1 M parameters.
Compared to the baseline~\cite{charoen2022rhode} with an accuracy of \qty{97.1}{\percent} and an F1-Score of \qty{97.13}{\percent}, the utilized model performs slightly worse with an accuracy of \qty{97.14}{\percent} and a F1-Score of \qty{90.89}{\percent}, however, requiring only 1 M instead of 56 M parameters.
Furthermore, the employed model needs drastically less parameters (1 M vs 195 M) compared to~\cite{abian2025atrous}, while outperforming the Swin transformer notably (F1-Score of \qty{90.89}{\percent} vs. \qty{69.85}{\percent}).
MobileNet already functions as a hardware-aware network which was specifically designed for small mobile devices.
For ultra-low power applications like VCE, this is still too complex and thus 1 M parameters would require a large area of a chosen hardware architecture.
We can further observe that the evaluation on RGGB images only marginally reduces classification performance.
Compared to RGB inputs, the accuracy is only reduced from $97.14\%$ to $96.20\%$ while precision and F1-Score decrease more significantly.
However, these findings indicate that employing a light-weight model directly on RGGB images as received from a camera still allows accurate image classification.

\begin{figure}[htbp]
    \centering
    \includegraphics[scale=0.5]{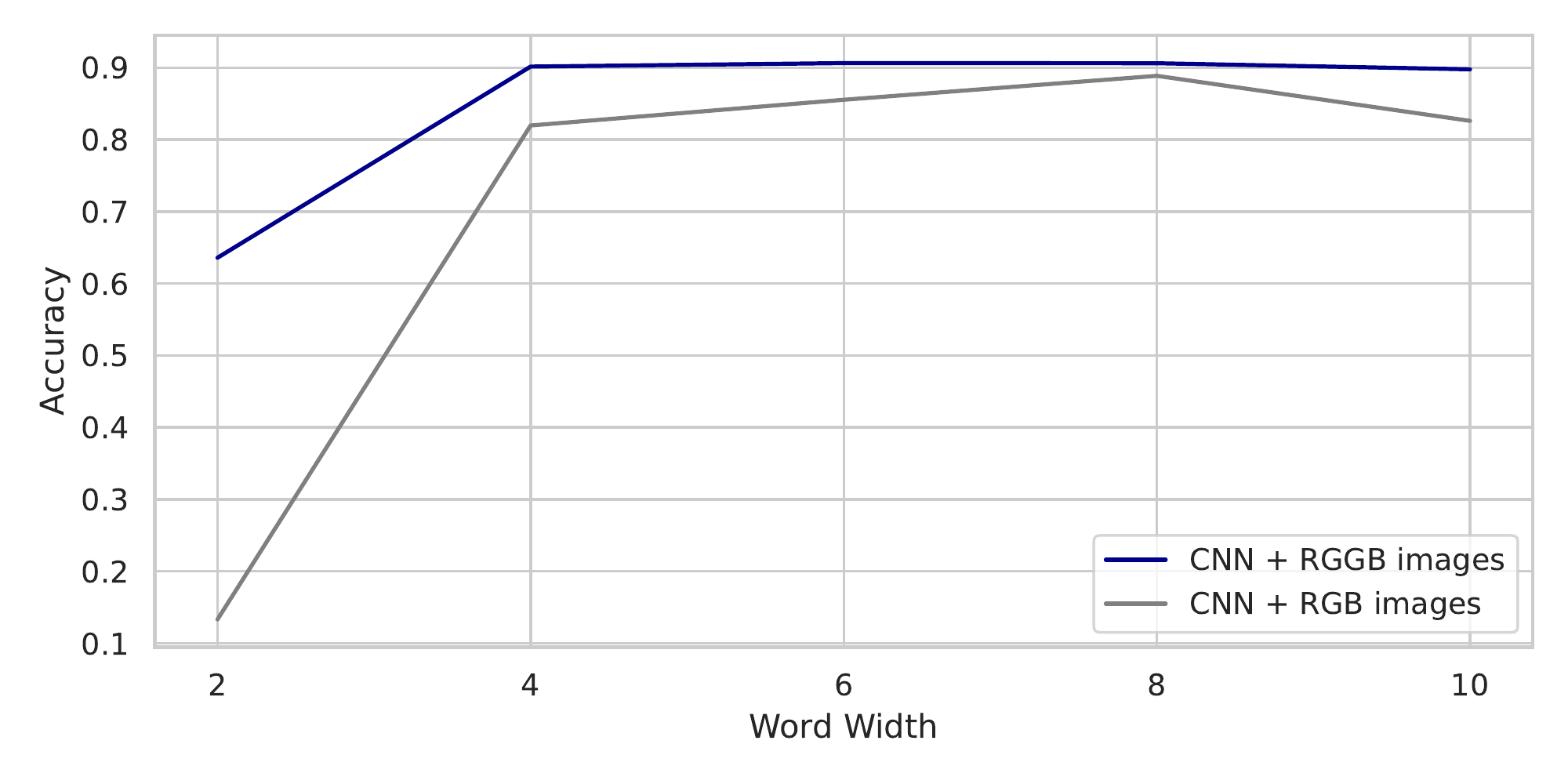}
    \caption{Classification performance of the hardware-suitable CNN over different word widths, validated either with raw Bayer pattern (RGGB) images or RGB images as input.}
    \label{fig:word_width}
\end{figure}

Thus, based on these results, we explored if an even less complex model with only $62,976$ parameters can also capture the class differences if RGGB images are received as an input in comparison to the standard RGB images.
Figure~\ref{fig:word_width} shows a comparison of the classification performance of the presented hardware-suitable CNN with RGGB images as input compared to standard RGB images over different word widths.

This demonstrates that this model is capable of classifying VCE images with a high accuracy even if directly executed on images in Bayer pattern format, without requiring a preceding conversion to RGB format.
Surprisingly, directly training on RGGB images seemingly performs better than using standard RGB images with this network.

\subsection{Postprocessing with Viterbi Decoding}
In the following, we aim to enhance the classification ability of this network even further by combining the CNN with an HMM and subsequent Viterbi decoding (see Table~\ref{tab:results_ut_net}).

\begin{table}[htbp]
	\caption{Classification performance of the hardware-suitable CNN (word width=$8$ bit) in combination with HMM and Viterbi Decoding (window size = $50$).}
            \begin{center}\resizebox{\columnwidth}{!}{
			\begin{tabular}{lccccc}
					\toprule
                     &\textbf{Accuracy} &  \textbf{Recall} & \textbf{Precision}  & \textbf{F1-Score} & \textbf{Params} \\
                    \midrule\midrule
                    Localization Net & 90.61& 83.98& 71.49 & 75.81 & 62,976\\
                    Localization Net \& HMM & 93.06 & 89.55 & 80.65 & 84.08 & 62,976\\
					\bottomrule
				\end{tabular}%
                }
\end{center}
\label{tab:results_ut_net}
\end{table}

\begin{figure}[hptb]
    \centering
    \includegraphics[width=\textwidth]{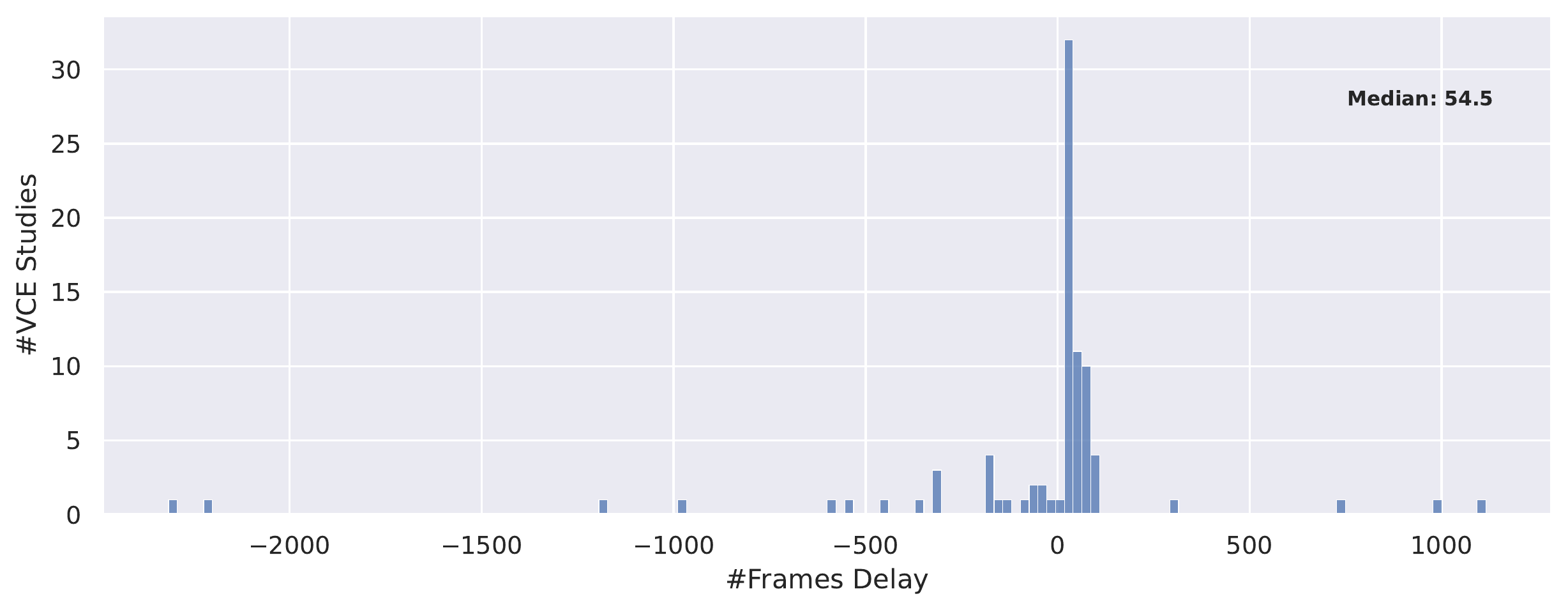}
    \caption{Distribution of delays across all VCE studies.}
    \label{fig:delays}
\end{figure}

Combining the localization net with an HMM enhances the classification performance to an accuracy of \qty{93.06}{\percent} and a sensitivity of \qty{89.55}{\percent}. 
While this model performs inferior compared to the baseline results, it still provides remarkable classification capabilities considering the low complexity.
The presented approach only requires $2,874,880$ MAC operations and a total of $62,976$ parameters, instead of 1 M.
Since during Viterbi decoding a number of observations need to be received to estimate the most likely point in time for observing a specific state (with the VCE as an example: the small intestine), employing this method can be accompanied by a certain delay.
Thus, for all patients within the test set the delay, meaning the difference between the actual first occurrence of the small intestine and the detection by the Viterbi decoding, is depicted in Figure~\ref{fig:delays}.

With a median delay of $54.5$, the plot visualizes that the HMM predicts the entering of the small intestine accurately for the majority of patients.
Only a few outlier studies resulted in extreme delays (either too early or too late detection), which are investigated closer in the following.

\begin{figure}[hptb]
		\centering
		\subfloat[Standard ST]{\includegraphics[width=0.2\linewidth]{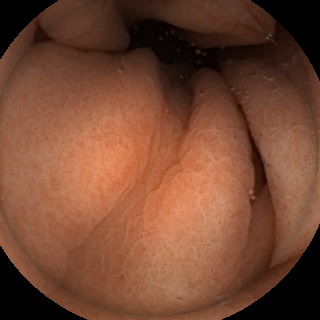}\label{fig:s298_st}}\hspace*{0.5mm}
		\subfloat[ID18-ST]{\includegraphics[width=0.2\linewidth]{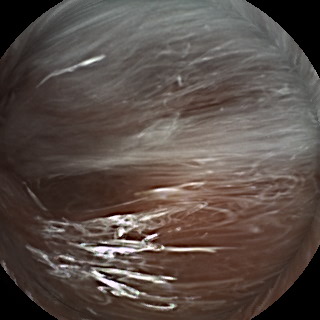}\label{fig:s018_st}}
        \hspace*{0.5mm}
		\subfloat[ID23-ST]{\includegraphics[width=0.2\linewidth]{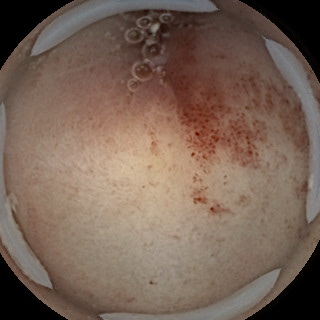}\label{fig:s023_st}}\hspace*{0.5mm}
        \subfloat[ID193-ST]{\includegraphics[width=0.2\linewidth]{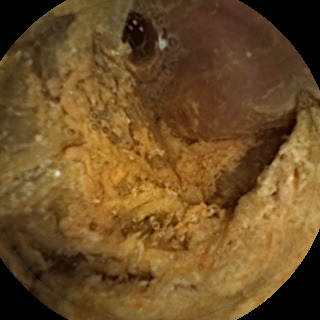}\label{fig:s193-st}}\\
        \subfloat[Standard SI]{\includegraphics[width=0.2\linewidth]{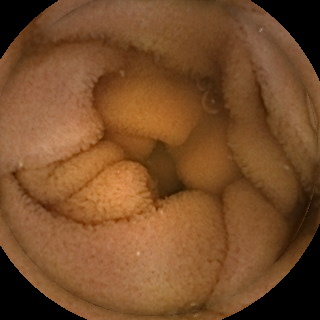}\label{fig:s298_si}}
		\hspace*{0.5mm}
		\subfloat[ID18-SI]{\includegraphics[width=0.2\linewidth]{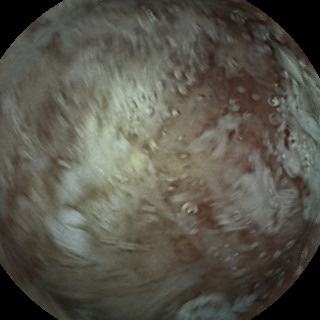}\label{fig:s018_si}}
        \hspace*{0.5mm}
        \subfloat[ID193-SI]{\includegraphics[width=0.2\linewidth]{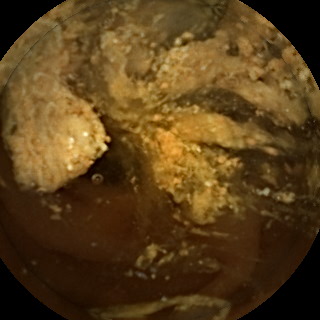}\label{fig:s193-si}}\hspace*{0.5mm}
		\subfloat[ID221-SI]{\includegraphics[width=0.2\linewidth]{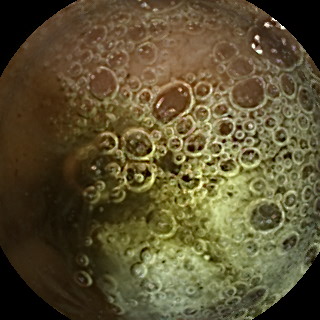}\label{fig:s221_si}}
		\caption{Examples of VCE images from the Rhode Island~\cite{charoen2022rhode} test set with frames originating from the stomach (ST) on top and frames from the small intestine (SI) on the bottom. (a) and (e) display a common image for each organ as present in the dataset. In comparison, (b)-(d) and (f)-(h) are from the studies resulting in the lowest accuracies or delays in the presented work.}
		\label{fig:vce-examples}%
	\end{figure}

In Figure~\ref{fig:vce-examples}, for the stomach as well as the small intestine, example VCE frames are shown with standard images for each organ in \ref{fig:s298_st} and \ref{fig:s298_si} as a reference and, in comparison, examples VCE images from the stomach \ref{fig:s018_st}-\ref{fig:s193-st} and from the small intestine \ref{fig:s018_si}-\ref{fig:s221_si} are displayed for outlier studies with the largest obtained delays.

The presented VCE example images display various obstacles such as bubbles and digestion remains.
The VCE study with the ID $23$ had a delay of $2180$, study $18$ a delay of $1915$.
For studies $221$ and $193$, the small intestine was detected $4680$ and $4496$ frames too early respectively.
The visual differences as shown in Figure~\ref{fig:vce-examples} seemingly challenge the image detection process.

\subsection{Demonstrator Performance}

As the battery capacity is constrained by the size of the capsule, the whole system cannot be active all the time transmitting 58 frames per second to the on-body receiver.
As shown in Table ~\ref{tab:results_power}, image capturing is the main energy consumer within the image processing pipeline.
Additionally, the transmission of images to the on-body receiver also requires its share of energy.

\begin{table}[htbp]
	\caption{Power (\unit{\milli\watt}) and energy (\unit{\micro\joule}) consumption by each module per frame processed with comparison to the AI capsule of~\cite{sahafi2022edge}. Metrics that are not mentioned in their paper are marked as undefined (U). Note that no energy consumption is listed for the idle state of our demonstrator, as the time in idle depends on the current frame rate, thus it is not constant.}
    \begin{center}\resizebox{\columnwidth}{!}{
    \begin{tabular}{l|ccc|c|c|cc}
            \toprule
            \textbf{Task} & \multicolumn{3}{c|}{Image Capture} & DNN & HMM & \multicolumn{2}{c}{Idle} \\
            \multirow{2}*{\textbf{Module}} & \textbf{Image} & \multirow{2}*{\textbf{LEDs}} & \multirow{2}*{\textbf{MCU}} & \textbf{MCU} & \multirow{2}*{\textbf{MCU}} & \multirow{2}*{\textbf{MCU}} & \textbf{Image}\\
            & \textbf{Sensor} & & & \textbf{with acc.} & & & \textbf{Sensor}\\
            \midrule\midrule
            Power Consumption & \multirow{2}*{$\sim60$} & \multirow{2}*{U} & \multirow{2}*{U} & \multirow{2}*{$\sim50$} & \multirow{2}*{-} & \multirow{2}{*}{N/A} & \multirow{2}{*}{0.1} \\
            of ~\cite{sahafi2022edge} [\unit{\milli\watt}] & & & & & & & \\
            \midrule
            Power Consumption [\unit{\milli\watt}] & 8.51 & 14.78 & 7.23 & 16.63 & 9.94 & 5.85 & 3.0\\
            Inference Time [\unit{\milli\s}] & 12.79 & 12.79 & 12.79 & 0.31 & 0.02 & - & -\\
            Energy Consumption [\unit{\micro\joule}] & 108.93 & 189.15 & 92.56 & 5.14 & 0.17 & - & - \\
            \bottomrule
        \end{tabular}%
        }
\end{center}
\label{tab:results_power}
\end{table}

The receiver typically draws \qty{5}{\milli\watt} and achieves \qty[per-symbol = p]{2}{\mebi\byte p\second} ~\cite{sahafi2022edge, liu2016design}.
An image has a size of \qty{102.4}{\kibi\byte}, thus requires \qty{50}{\milli\s} to be transmitted to the on-body receiver.
The resulting energy consumed per frame is \qty{\sim 250}{\micro\joule}.
Thus, reducing the amount of worthless images captured and transmitted increases the battery life of the VC.
By incorporating UltraTrail and a quantized implementation of the HMM, the captured images can be analyzed to determine the current location within the GI tract.
The energy consumed to process the images is at \qty{0.31}{\micro\joule} negligible compared to the image sensor, the LEDs, and the transmitter module. 
Thus, the firmware captures images with a defined frame rate and analyzes them with the HW accelerator in combination with Viterbi decoding. 
The capsule's arrival in the small intestine, as indicated by the HMM, initiates the actual screening process thereby increasing the frame rate and triggering the subsequent transmission of image data.

\subsubsection{Simulation of GI Tract Traversals:}
To verify the functionality of our demonstrator and evaluate its performance, multiple studies from the Rhode Island test set are simulated by the HIL setup~\cite{bause2025hil}.
The studies were captured by using the PillCamTM SB3 capsule, which utilizes a similar image sensor with a resolution of $320\times320$ pixels and a dynamic frame rate of 2 to 6 fps.
In order to ensure the comparability of different simulations, it is assumed that the frame rate is fixed at 2 fps during the study's original recording.
The objective of on-edge location detection is to identify the point of entering the area of interest, the small intestine, with minimal delay while simultaneously spending only as much energy as absolutely necessary.
A delay that is less than zero indicates that the transmission is initiated prematurely. 
This results in the consumption of energy that could be allocated for more crucial images. 
Conversely, a high delay suggests that the transmission is initiated after the optimal time, resulting in the skipping of segments at the beginning of the small intestine.
As the standard endoscopic examination of the gastroscopy is capable of screening up to \qty{30}{\centi\metre} of the small intestine~\cite{Lewis2015}, a slight positive delay is permissible.
The baseline of the experiment is an ordinary video capsule, that captures 2 fps and transmits them without further processing to the on-body receiver.
Thus, the baseline has a deeply negative delay as it sends all frames, also those of the esophagus and stomach.

\begin{figure}[hptb]
    \centering
    \includegraphics[width=\textwidth]{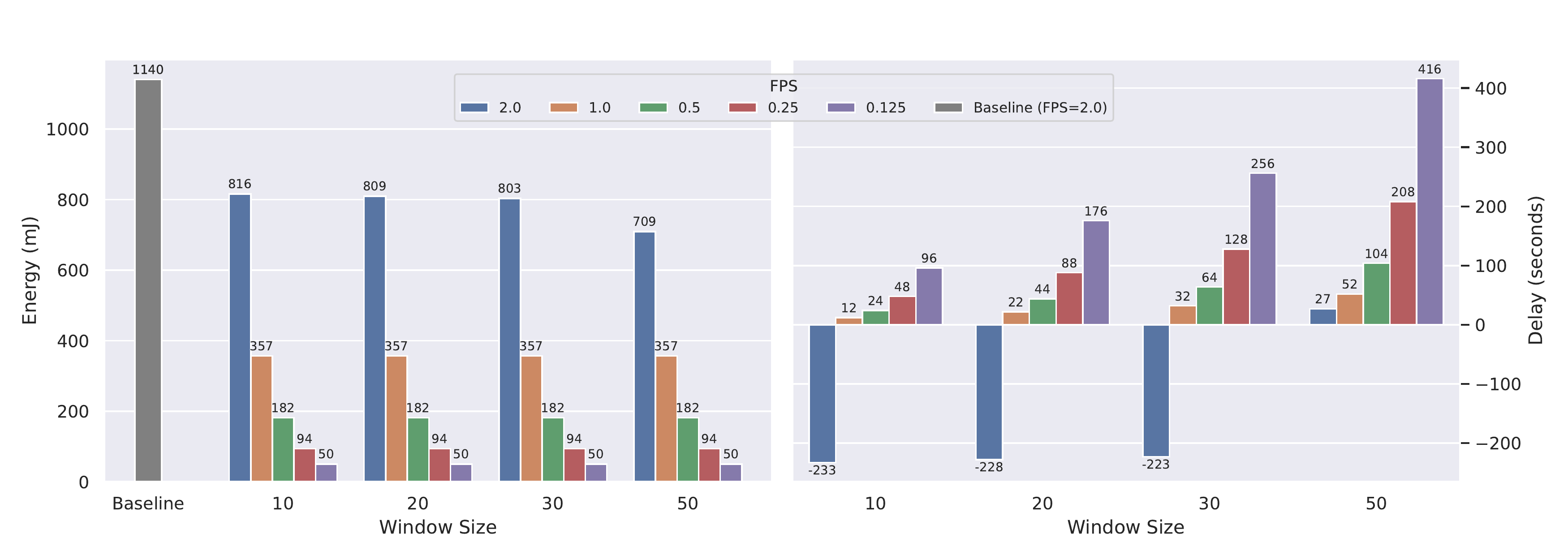}
    \caption{Energy consumption (left) and delay in frames (right) of different HMM window size and fps combinations for the simulation of study 73.}
    \label{fig:energy}
\end{figure}

To illustrate the capabilities of the location detection, study 73 of the Rhode Island test set is used as exemplary screening.
As shown in Figure ~\ref{fig:energy}, the baseline capsule spends \qty{1140}{\milli\joule} of energy prior to attaining the area of interest, which is effectively dissipated.
Using the same frame rate and analyzing the images on-edge, however, only requires \qtyrange[range-units = single]{709}{816}{\milli\joule} even though a small HMM window size leads to a premature misdetection of the small intestine, which starts the image transmission too early.

\begin{figure}[hptb]
    \centering
    \includegraphics[scale=0.35]{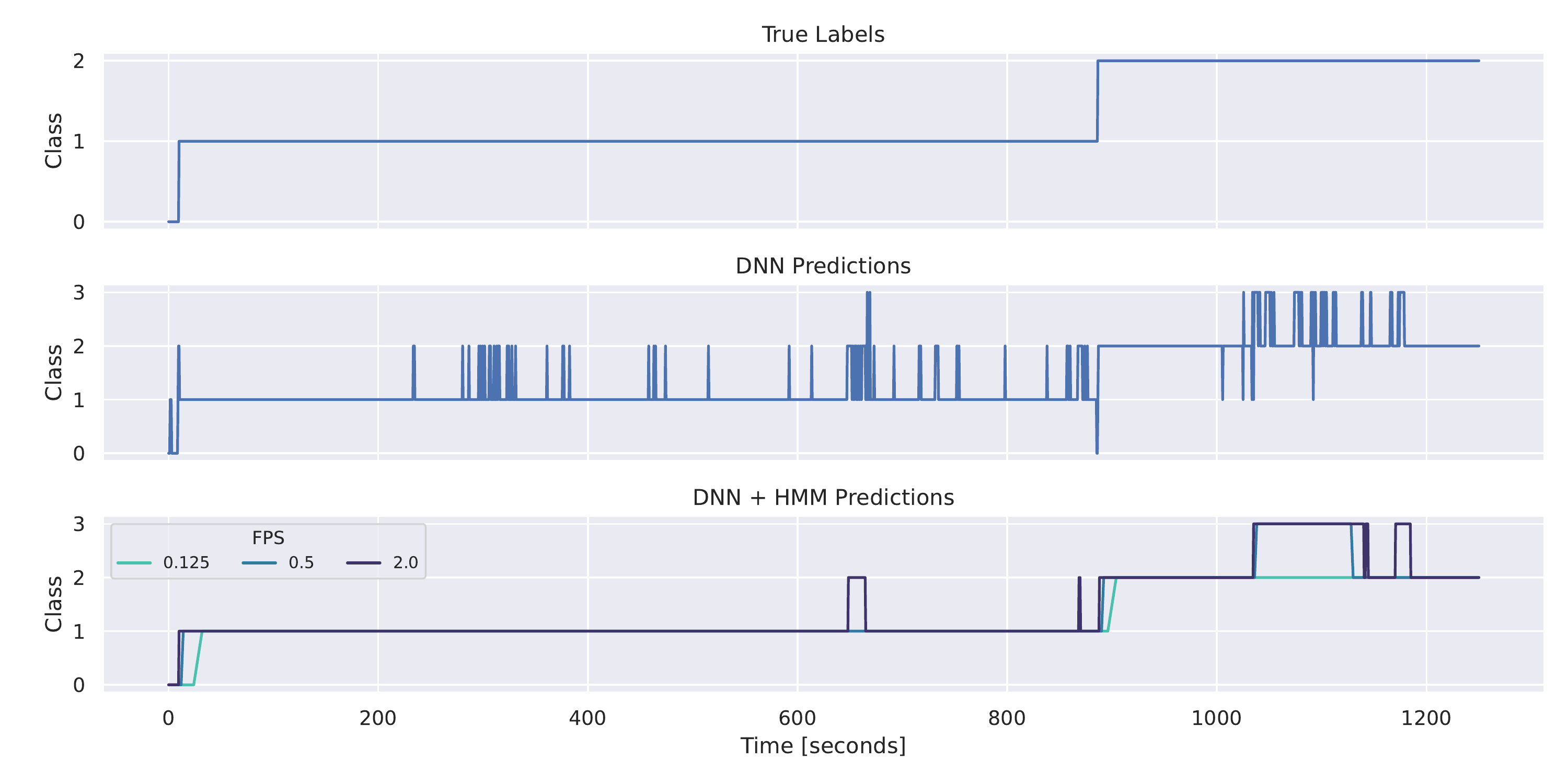}
    \caption{Original labels, DNN predictions, and HMM prediction of different fps with a window size of 10 for study 73 (with the classes: 0 - Esophagus, 1 - Stomach, 2 - Small intestine, 3 - Colon).}
    \label{fig:preds}
\end{figure}

This misdetection is initiated by a series of successive images of compromised quality (Figure ~\ref{fig:preds}, DNN Predictions), which collectively challenge the capabilities of the CNN.
Therefore, a decrease in the frame rate has two primary benefits. 
It reduces energy consumption and it also decreases the likelihood of consecutive frames containing dirt.

\begin{figure}[hptb]
    \centering
    \subfloat[Average Energy Consumption \lbrack\unit{\milli\joule}\rbrack\\\hspace*{1.55em}(avg. of baseline: \qty{719.934}{\milli\joule})]{\includegraphics[width=0.5\textwidth]{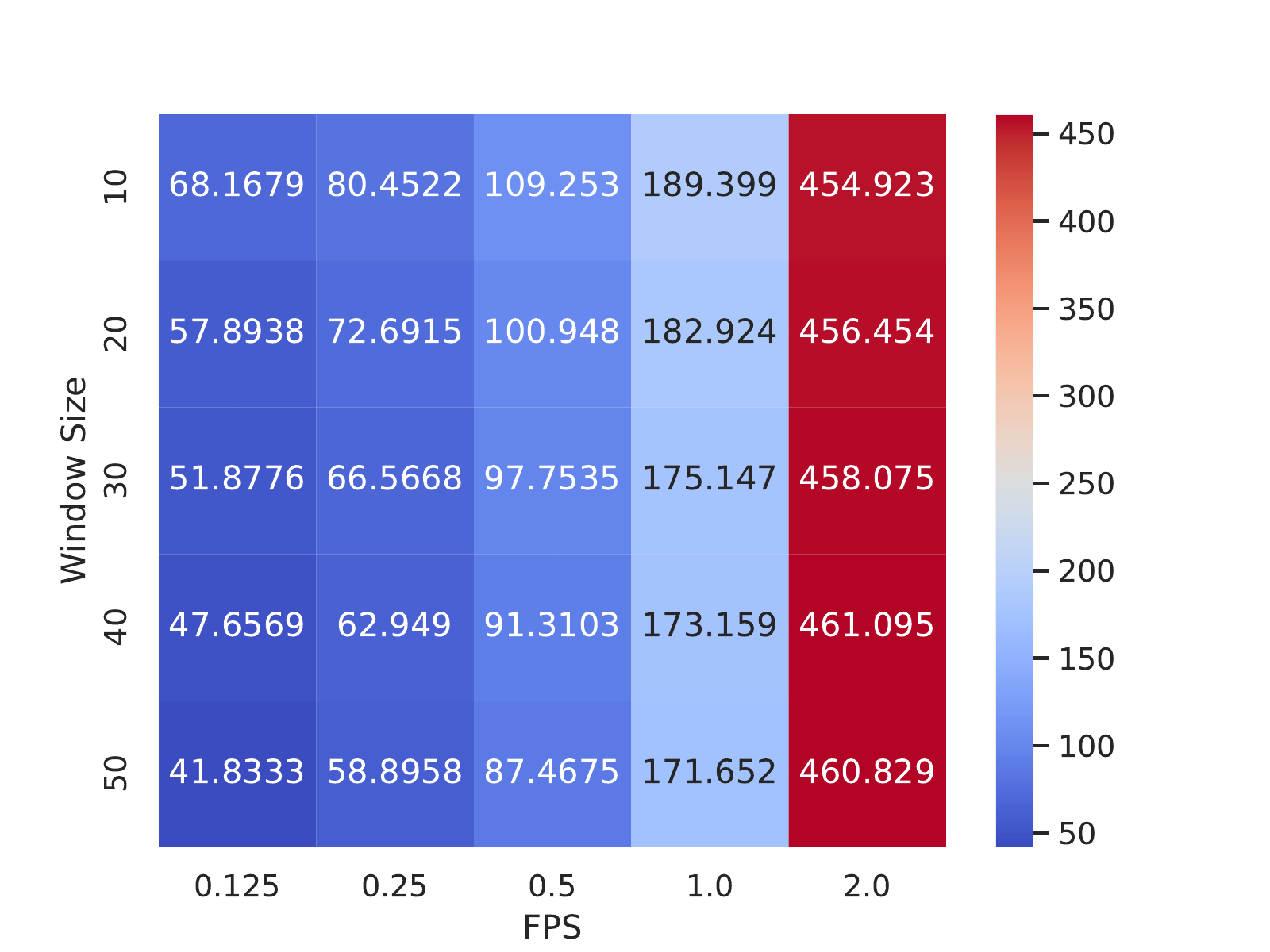}\label{fig:avg_energy}}
    \subfloat[Average Delay \lbrack\unit{\s}\rbrack]{\includegraphics[width=0.5\textwidth]{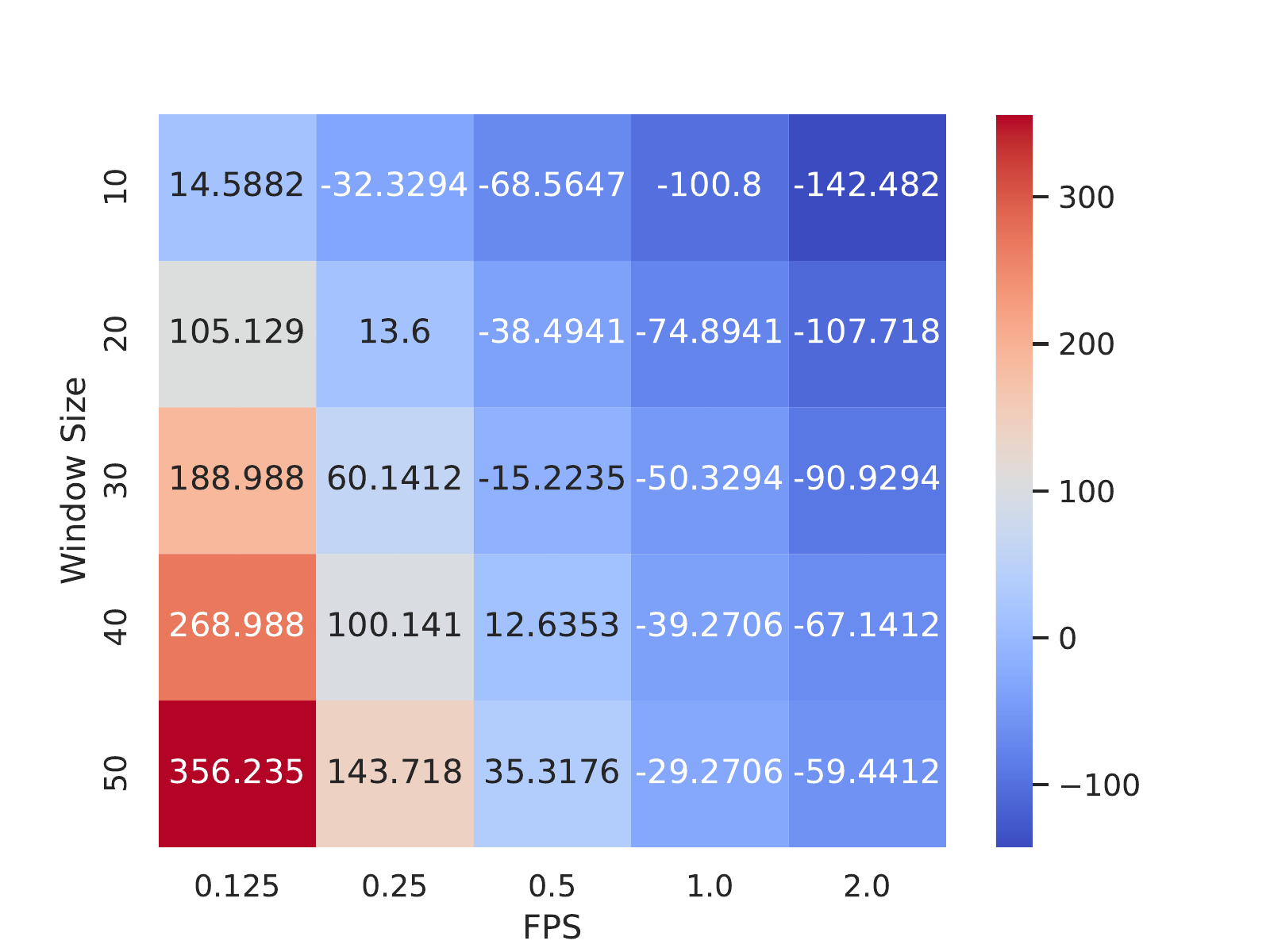}\label{fig:avg_delay}}
    \caption{Grid search of the HMM hyperparameters window size and frame rate to find the optimal combination in regard of average energy consumption (a) and average delay (b) across all test studies of the Rhode Island data set.}
    \label{fig:grid_search}
\end{figure}

It should be noted that this process concomitantly engenders an increase in the delay between the transition into the small intestine and the HMM detecting the transition as more time elapses before the window is filled with a path pointing towards a small intestine prediction.
Thus, a trade-off between frame rate and HMM window size needs to be found to achieve a low energy consumption while keeping the delay close to zero.
Figure~\ref{fig:grid_search} displays the results of a hyperparameter grid search that was performed on all Rhode Island test studies.
The baseline achieved an average energy consumption across all studies of \qty{719.934}{\milli\joule}.

As expected, the average energy consumed increases with a higher frame rate.
For example, an increase in frame rate from 1 to 2 fps multiplies the energy consumed by $\sim 240\%$ since a higher frame rate can result in an early detection of the small intestine, because of multiple consecutive images of a same scene that challenges the CNN, and thus starts the transmission prematurely.
On the other hand, a very low frame rate combined with a large window size is energy-saving but it inserts a massive average delay of more than \qty{350}{\s} in which the capsule may travel further through the small intestine than the \qty{30}{\centi\metre} that can be screened using the standard endoscopic examination. 
As a result, crucial segments of the small intestine may not be screened at all.
A frame rate of $0.25$ fps with a window size of $20$ seems to be a good trade-off between energy consumption and delay and saves $\sim 89.9\%$ energy compared to the baseline capsule.
If more energy is available, $0.5$ fps with a window size of $30$ or even $40$ could also be possible to make post-processing more stable against some mispredictions of the CNN.

\section{Conclusion}
In the presented work, extensive simulation from capturing images with a Nan-EyeC miniature camera sensor to final evaluation with a hardware suitable approach composing of a light-weight CNN and subsequent quantized Viterbi decoding is shown.
By training the model directly on raw Bayer images, an additional costly conversion step implementation on the system's architecture is avoided.
While requiring only $62,976$ parameters for the CNN, an accuracy of \qty{93.06}{\percent} is achieved.
The power and energy consumption for each module is reported for this setup, yielding a total energy consumption of only \qty{5.14}{\micro\joule} for the DNN and \qty{0.17}{\micro\joule} for the Viterbi decoding per frame analyzed.
A grid search demonstrated that it is advisable to decrease the frame rate to $0.25$ fps with a HMM window size of $20$ to achieve a $89.9\%$ lower power consumption while still detecting the transition into the small intestine timely.
It is demonstrated that the presented setup is capable of real-time on-edge location detection.

 \bibliographystyle{splncs04}
 \bibliography{bibliography}

\end{document}